# Doubly Topological Harmonic Generation

Kaushik Kudtarkar[1], Xinyi Wang[2], Yunjo Jeong[1], Chloe F. Doiron[3], Alexander Cerjan[3*], and Shoufeng Lan[1,2,4*]

**Affiliations:**
[1]Department of Mechanical Engineering, Texas A&M University, College Station, TX, USA.
[2]Department of Material Science and Engineering, Texas A&M University, College Station, TX, USA.
[3]Center for Integrated Nanotechnologies, Sandia National Laboratories, Albuquerque, NM, 87185, USA.
[4]Department of Electrical & Computer Engineering, Texas A&M University, College Station, TX 77843, USA
*Corresponding Authors: shoufeng@tamu.edu and awcerja@sandia.gov

**Abstract:** The proposition that band geometry alone can protect optical states against disorder has proven not merely theoretically elegant but experimentally incontrovertible. A key attribute of photonic topological systems is their capacity to simultaneously possess high-intensity excitations at multiple distinct frequencies that are confined to the same topological interface. However, exploiting this freedom to protect the interaction between at least two topological states has remained an open experimental challenge. Here, we report an interaction between two topological states, with one being precisely frequency-doubled to the other, supported in a hybrid plasmonic and photonic topological insulator via nonlinear phase matching. We find that the phase matching inherits a unique spin-momentum locking unseen in conventional nonlinear systems. This ability to bring two topological states into phase-matched nonlinear interaction at a single interface sets the stage for a new class of doubly protected nonlinear photonic devices, potentially finding implications in generating entangled photon pairs with enhanced resilience and robustness for secure quantum information technology.

**Main Text:**

## Introduction

Topological photonics has paved a new pathway to manipulate light in a robust manner against scattering losses caused by defects and disorder (*1*). The past decade has witnessed the surging development of this research field (*2, 3*). Inspired by the dissipationless edge conductance of electrons in the integer quantum Hall effect and its connection to topology (*4-8*), early research in topological photonics focused on backscattering-free boundary states of magneto-optical systems with broken time-reversal symmetry (*9-11*). Alternative methods to circumvent the weak magnetic responses in high-frequency optical regimes exploit synthetic magnetic fields and index-modulation in an extra dimension for breaking the time-reversal symmetry (*12, 13*). Analogous to quantum spin and valley Hall (QSH and QVH) effects, time-reversal symmetry invariant photonic systems can achieve robust boundary-localized states without a magnetic or synthetic magnetic field (*14*); these systems include coupled-resonator optical waveguides (*15, 16*), photonic crystals (*17-20*), and quasicrystals (*21, 22*). The key is bulk-edge correspondence, in which the two bulk systems forming the edge have different topologies.

While topological photonics has experienced a burgeoning development by emulating condensed matter physics, future advancement would go beyond direct emulation, which will rely on the distinctions of photonic systems, such as the ability to access multiple energy levels simultaneously for nonlinear optical responses (*23*). Indeed, nonlinear responses have been shown to induce topological phase transitions (*24, 25*), edge solitons in topological bandgap (*26-29*), and third-harmonic generation in nanoscale resonator arrays (*30*). While optical nonlinearity was present in these demonstrations, they rely more or less on a single topological state. For interacting two states simultaneously, sum-frequency generation was used to drive nonlinear photonic crystals periodically (*31*). However, the induced topological band gap was only observed at a single frequency. Recently, a theory predicted that second-harmonic generation (SHG) in QVH photonic crystals could couple double topological states with their interactions evident in nonlinear phase matching (*32*).

Here, we experimentally observe SHG between two boundary-localized states that are both protected by the QVH effect in a hybrid plasmonic and photonic topological insulator (HPPTI) (*33*). By engineering the topological edge between two silicon nitride ($Si_3N_4$) domains with inverted valley topology, we realize topological edge modes (EMs) at both fundamental ($\lambda_{FF}$ = 900 nm) and second harmonic frequency ($\lambda_{SH}$ = 450 nm) that share similar spatial profiles and satisfy a phase-matching condition. The silver-backed reflector enhances vertical field confinement within the active Si3N4 layer, yielding a threefold increase in SHG intensity compared to a dielectric-only structure. Helicity-resolved measurements at each frequency independently confirm spin-momentum locking. More importantly, phase-matching-dependent SHG spectra further show that the nonlinear emission inherits the helicity of the topological EM, enabling directional control of the second harmonic output via the pump polarization.

## Result and discussions

Realizing efficient on-chip SHG requires simultaneously confining light at both the fundamental frequency ω and the second harmonic 2ω to the same spatial channel while satisfying momentum conservation. Conventional photonic waveguides struggle to meet these conditions simultaneously because the modes at the two frequencies occupy distinct spatial regions within the structure, thereby reducing their overlap. Since achieving momentum conservation between them requires careful dispersion engineering that is difficult to maintain across fabrication variations, we address both challenges by using a QVH HPPTI formed from two topologies with opposite Berry curvatures. The topological domains consist of $Si_3N_4$ cylinders of radii $r_1 = 0.13a$ and $r_2 = 0.28a$ with a lattice constant a = 370 nm, on a 25 nm silicon dioxide ($SiO_2$) spacer above a 100nm silver back reflector (refer supplementary notes 1 and 2), and together they generate a topologically protected EM at their boundary (Fig. 1A). Because the bulk regions on either side remain gapped at the frequencies of interest, meaning they cannot support freely propagating optical states, light injected at the topological edge has no pathway into the bulk and instead couples entirely into the edge channel, propagating along the boundary without leaking into the surrounding lattice. Real-space imaging confirms edge-guided transport at both wavelengths of $\lambda_{FF} = 900$ nm and $\lambda_{SH} = 450$ nm (Fig. 1B) (refer supplementary note 8). Excitation at wavelengths outside these EMs instead produces delocalized bulk modes that spread across the lattice, directly mapping the spectral boundaries of the topological edge states and confirming that the confinement is a result of the band topology rather than a geometric resonance.

The ribbon band structure (Fig. 1C) reveals edge-localized modes embedded within the bulk bandgaps at both frequencies. These modes share similar transverse spatial profiles (refer supplementary note 3), meaning that the optical fields at $\omega$ and $2\omega$ are concentrated in the same physical region of the topological edge. This spatial co-localization is a prerequisite for strong nonlinear interaction, since the efficiency of SHG depends directly on the integral of the product of the two field distributions across the waveguide. A large overlap integral indicates that the nonlinear polarization driven by the fundamental field efficiently radiates into the second harmonic mode, rather than being lost to modes with incompatible spatial profiles. Crucially, the doubled fundamental dispersion $2*(\omega, k_{\omega})$ intersects the second harmonic dispersion $(2\omega, k_{2\omega})$ at a finite wavevector away from the Brillouin zone center (Fig. 1D), satisfying both energy and momentum conservation at a well-defined off-$\Gamma$ phase matching point. This intersection arises because the engineered dispersion of the EMs allows their slopes and curvatures to align at a finite wavevector, a condition that persists when $r_1$ is varied by 0.01a and that shifts both dispersions while preserving multiple intersection points. This robustness against geometric variation distinguishes the present approach from conventional quasi-phase-matched schemes, where matching conditions are sensitive to fabrication tolerances and must be maintained over the full length of the device.

A scanning electron microscope image of the fabricated device (refer supplementary note 6) shows the topological edge clearly separating two trivial domains (Fig. 2A). To experimentally distinguish the optical response of the bulk from that of the topological edge, we collect reflection spectra from localized regions within the trivial domains and directly at the topological edge (Fig. 2B). Using reflection spectroscopy (refer supplementary note 7), EMs appear as dispersive features whose spectral positions shift with the angle of incidence, corresponding to a change in the in-plane wavevector of the probing light. The trivial bulk region is featureless within the bandgap, as expected, because no optical states exist there to couple to the incident field. In contrast, spectra acquired at the topological edge show pronounced dispersive features at both fundamental ($\lambda_{FF} =$ 900 nm ) and second harmonic frequency ($\lambda_{SH} = 450$ nm), in quantitative agreement with simulation. The momentum-dependent evolution of these features with in-plane wavevector confirms that they arise from guided topological edge states and not from localized defect modes, which would appear as flat, non-dispersive features in the same measurement. A defining property of QVH EMs is spin-momentum locking. The valley degree of freedom in the photonic crystal corresponds to distinct momentum-space locations where the bands carry concentrated Berry curvature of a given sign. The optical field carries spin angular momentum determined by its circular polarization state, and this spin couples to the valley index through the symmetry of the crystal, so the helicity of the incident light selects which valley is excited and therefore which direction the EM propagates. Spin-momentum locking at both frequencies is demonstrated experimentally (Fig. 2C) . At the fundamental wavelength ($\lambda_{FF}$ = 900 nm), left circularly polarized (LCP) and right circularly polarized (RCP) excitation launch counter-propagating EMs, with the propagation direction reversing upon switching helicity, consistent with simulations. Similarly, at the second harmonic frequency, LCP and RCP excitation at $\lambda_{SH}$ = 450 nm launch counter-propagating EMs at $\lambda_{SH}$ = 450 nm, again with the direction reversing upon switching helicity. The fact that spin-momentum locking is independently observed at both frequencies confirms that each EM individually exhibits chiral transport governed by the valley topology of the interface.

Silicon nitride is conventionally centrosymmetric with no bulk $\chi^{(2)}$, but PECVD deposition induces silicon nanocluster formation that locally breaks centrosymmetry, giving rise to a strong bulk-type second-order nonlinearity confirmed by its quadratic thickness dependence (*34-36*). The deposition-induced bulk $\chi^{(2)}$ of silicon nitride serves as the nonlinear engine of the platform presented here, where a QVH photonic crystal interface confines topological edge modes at both the fundamental and second harmonic frequencies. However, in dielectric photonic crystals, optical fields can leak vertically into the substrate because the refractive index contrast between the patterned layer and the underlying material is insufficient to provide complete vertical confinement (*37*). This leakage reduces the local field intensity within the nonlinear medium and suppresses conversion efficiency, since the fraction of optical energy that escapes into the substrate does not contribute to the nonlinear interaction. We mitigate this by introducing a 100 nm plasmonic silver back reflector separated from the $Si_3N_4$ layer by a 25 nm $SiO_2$ spacer (refer supplementary note 4). The $SiO_2$ spacer provides controlled separation from the reflector to prevent excessive absorption losses. The reflector redirects the optical field back into the active region before it dissipates into the substrate, thereby increasing the field amplitude at the nonlinear interface. Simulations of the out-of-plane electric field magnitude $|E_z|$ at $\lambda_{FF}$ = 900 nm confirm this picture (Fig. 3A). The bare dielectric structure exhibits substantial field penetration into the substrate, whereas the metal-backed stack strongly confines the field within the $Si_3N_4$ layer.

Since SHG intensity scales quadratically with fundamental power, $I(2\omega) \propto [I(\omega)]^2$, even a moderate improvement in field confinement translates into a disproportionately large gain in second harmonic output. This effect is quantified through log-log plots of SHG intensity versus input power for both structures (Fig. 3B). Both yield slopes are consistent with second-order nonlinear scaling, measured at 1.98 with the silver layer and 2.02 without, confirming that the frequency doubling mechanism is unchanged by the introduction of the metal layer and that no higher-order nonlinear processes contribute significantly. The silver-backed device produces consistently higher SHG intensity across the full power range, with the hybrid metal-dielectric platform delivering approximately a threefold enhancement over the dielectric-only structure. This enhancement persists across all measured power levels, confirming that it does not arise from a threshold effect or a resonance associated with the silver backing, but rather from the systematic increase in field confinement that the back reflector provides. The inset spectrum shows a sharp, isolated peak at $\lambda_{SH}$ = 450 nm with no competing spectral features, confirming spectrally clean frequency doubling under phase-matching conditions. This enhanced SHG signal remains topologically guided as observed experimentally (Fig. 3C). Under circularly polarized excitation at 900 nm, the second harmonic emission at 450 nm stays confined to the triangular topological edge and follows its geometry faithfully through sharp bends. LCP and RCP inputs launch SHG propagation in opposite directions, confirming that the silver layer enhances the nonlinear output without disturbing the chiral transport properties of the underlying topological EM.

To isolate the role of phase matching in governing the nonlinear response, we tune the fundamental wavelength from 885 nm to 915 nm and record the SHG intensity as a function of both the input and emission wavelengths (Fig. 4A) (refer supplementary note 9). The SHG peak shifts systematically with input wavelength at a rate consistent with exact frequency doubling, tracking the dispersion of the second harmonic EM. It reaches a pronounced maximum near $\lambda_{FF}$ = 900 nm, precisely where the calculated phase matching condition between the fundamental and second harmonic EMs is satisfied. Away from this condition, the nonlinear polarization generated at ω accumulates a growing phase mismatch with respect to the freely propagating 2ω field as both waves travel along the edge. This mismatch causes contributions generated at successive points along the waveguide to interfere destructively, suppressing coherent buildup and reducing the output substantially. This wavelength dependence therefore confirms that the observed SHG enhancement arises from constructive phase accumulation along the topological edge, rather than from a single-pass or background nonlinear process that would show no such spectral selectivity. Real-space SHG images at selected emission wavelengths from 440 nm to 460 nm (Fig. 4B), corroborate this interpretation. At $\lambda_{SH}$ = 450 nm, where phase matching is satisfied, the emission is sharply confined to the triangular topological edge and faithfully traces its geometry, reflecting the fact that the coherently built-up second harmonic field is guided by the same topological boundary that confines the fundamental. At off-resonant wavelengths, the signal weakens and spreads into the bulk. This spatial delocalization at off-resonant wavelengths shows that SHG is generated throughout the illuminated area. However, only the edge-guided phase-matched contribution builds coherently to produce a spatially and spectrally localized output. Phase matching therefore, acts simultaneously as a spectral filter and a spatial selector, concentrating the nonlinear emission into the topological channel where it can be efficiently collected and routed.

The direct experimental connection between topology, helicity, and nonlinear phase matching is established through the measurements presented (Fig. 5). Reflection spectra recorded under RCP excitation (Fig. 5A), and LCP excitation (Fig. 5B), at a fundamental wavelength of $\lambda_{FF}$ = 900 nm with the detector tuned to the SHG band, reveal localized intensity features in the wavelength and wavevector plane. Their positions coincide with the calculated intersections of $2\cdot(\omega, k_{\omega})$ and $(2\omega, k_{2\omega})$, confirming that the enhanced SHG signal occurs precisely at the momentum states where both energy and momentum conservation are simultaneously satisfied by the phase-matched topological EMs. The localization of these features in wavevector space, rather than their appearance as a broad background, is a direct signature of phase-matched nonlinear generation at a dispersive guided mode rather than at a bulk or surface state. Because LCP and RCP couple to opposite valley momenta, they access different sides of the phase matching condition in reciprocal space, shifting the wavevector at which the enhanced SHG appears. A line cut at the phase-matching wavevector (Fig. 5C) makes this explicit as RCP and LCP excitation produce asymmetric peaks at opposite values of wavevector $k_x$, directly confirming that the nonlinear response is direction-dependent and inherits the helicity of the topological EM. Taken together, we establish a consistent and complete experimental picture in which the QVH interface simultaneously localizes light at ω and 2ω, satisfies phase matching between the two EMs, and enables helicity-controlled directional switching of the nonlinear emission through spin-momentum locking, a combination that has no analog in conventional waveguide geometries.

## Conclusion

We have demonstrated that two distinct topological states at drastically different frequencies can be co-localized at a single QVH interface and coupled through nonlinear phase matching in a judiciously designed HPPTI. By engineering the HPPTI, we simultaneously obtain two topological edge modes where one frequency is precisely double the other. The two topological states can interact with each other through SHG enabled by induced $\chi^{(2)}$ nonlinearity in $Si_3N_4$ via PECVD. The $\chi^{(2)}$ responses are further enhanced and confined by a plasmonic silver reflector and a $SiO_2$ spacer to provide the required refractive index contrast. More importantly, we observe phase matching with spin-momentum locking, which ties the propagation direction of the nonlinear emission to the pump's helicity, a property without analog in conventional SHG. As spontaneous parametric down-conversion (SPDC) is the quantum counterpart of SHG, it would be fascinating to explore whether the same principle of co-localizing topological states at phase-matched frequencies can be extended. Further, because SPDC plays a central role in generating entangled photon pairs, this work might inspire a new route toward secure quantum information technology with enhanced resilience.

**Acknowledgments:** The authors gratefully acknowledge the funding support provided by Texas A&M University, United States National Science Foundation (Grant No. 2348611 and 2348610), and Sandia National Laboratories (Grant No. PO 2543653/1923579) from the United States Department of Energy (DOE). The nanofabrication was conducted in the Texas A&M University AggieFab Nanofabrication Facility (RRID: SCR_023639), which is supported by the Texas A&M Engineering Experiment Station and Texas A&M University. A.C. acknowledges support from the Laboratory Directed Research and Development program at Sandia National Laboratories. This work was performed in part at the Center for Integrated Nanotechnologies, an Office of Science User Facility operated for the U.S. Department of Energy (DOE) Office of Science. Sandia National Laboratories is a multimission laboratory managed and operated by National Technology & Engineering Solutions of Sandia, LLC, a wholly owned subsidiary of Honeywell International, Inc., for the U.S. DOE's National Nuclear Security Administration under Contract No. DE-NA-0003525. The views expressed in the article do not necessarily represent the views of the U.S. DOE or the United States Government.

**Author contributions:**

Conceptualization: SL, KK, AC

Methodology: SL, KK, AC, XW

Investigation: SL, KK, AC, XW, YJ, CD

Visualization: SL, KK

Funding acquisition: SL, AC

Project administration: SL, AC

Supervision: SL, AC

Writing – original draft: SL, KK, AC, XW, YJ, CD

Writing – review & editing: SL, KK, AC, XW, YJ, CD

**Competing interests:** Authors declare that they have no competing interests.

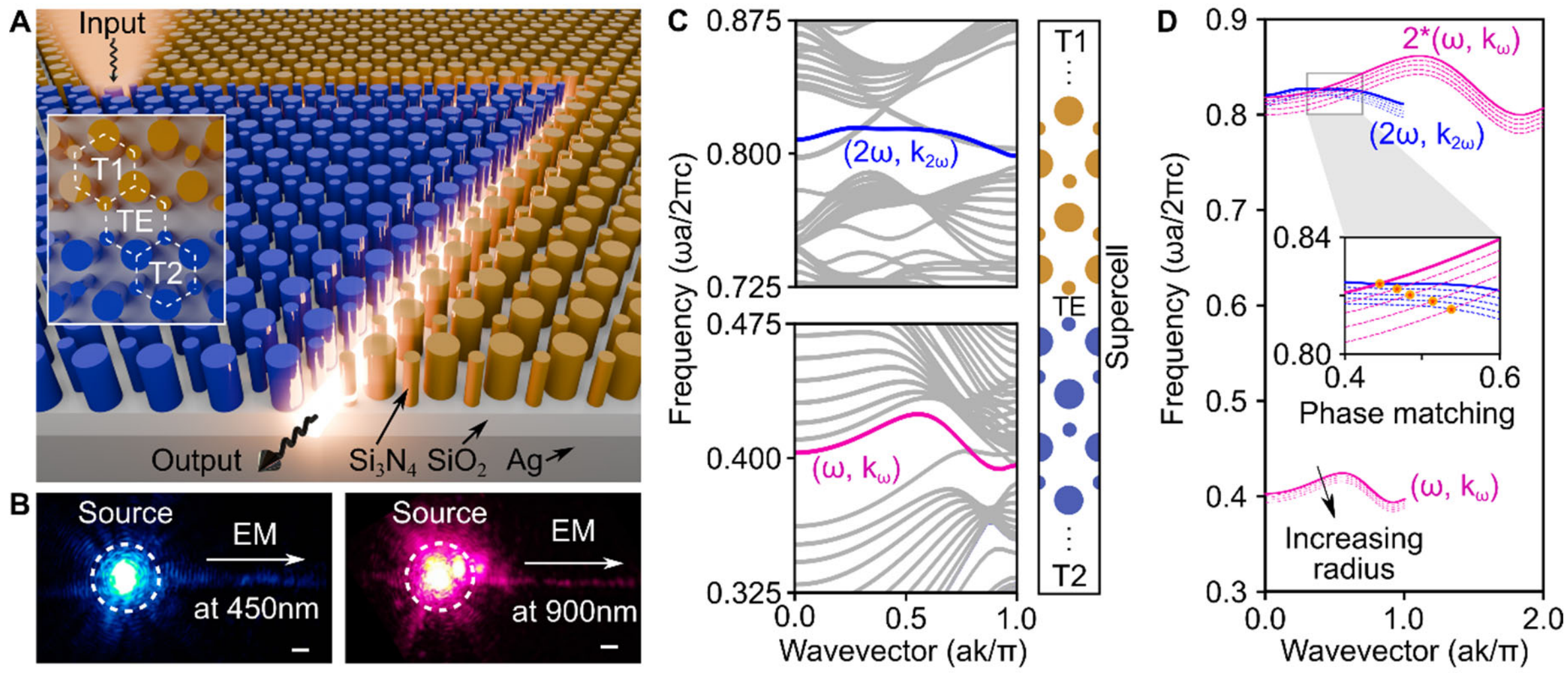

**Fig. 1. Double topological states interacting through nonlinear phase matching.** **(A)** Schematic (not to scale) of a hybrid plasmonic and photonic topological insulator (HPPTI) for supporting two topological states simultaneously and their interaction via nonlinear phase matching. Two domains (inset) with inverted lattices and associated topologies of T1 (blue) and T2 (gold) form a topological edge (TE) that can support topologically protected edge modes (EMs). Specifically, it is achieved through the quantum valley Hall effect in hexagonal lattices that consist of $Si_3N_4$ cylinders with alternating radii. Additionally, a $SiO_2$ spacer and a plasmonic silver back reflector are designed to enhance and confine EMs to the nonlinear $Si_3N_4$ cylinders, where a second-order nonlinearity ($\chi^{(2)}$) is purposefully induced during plasma-enhanced chemical vapor deposition (PECVD). **(B)** Two topological EMs propagating along the topological interface (indicated by the arrow) at fundamental and second harmonic wavelengths of 450 nm and 900 nm are simultaneously supported and experimentally observed in the HPPTI. Scale bars represent 5 µm. **(C)** Theoretically calculated results using a supercell (schematic: two topologies, T1 and T2, forming a topological edge, TE) reveal the two topological EMs, ($\omega$, $k_\omega$) and ($2\omega$, $k_{2\omega}$), formed within the bulk bandgap at both the fundamental (pink) and second harmonic (blue) frequencies. **(D)** The two topological modes are further interconnected via nonlinear phase matching, which is achieved by innovatively transforming the band at the fundamental frequency (pink) through doubling both frequency and wavevector, $(\omega, k_\omega)\rightarrow 2(\omega, k_\omega)$. This band transformation accurately describes the second harmonic generation (SHG) process, and the phase matching ($k_{2\omega} = 2k_\omega$) is obtained (inset) at the crosspoints (stars) between the transformed band and that of ($2\omega$, $k_{2\omega}$). This phase matching is controlled by tuning the radius of the cylinders.

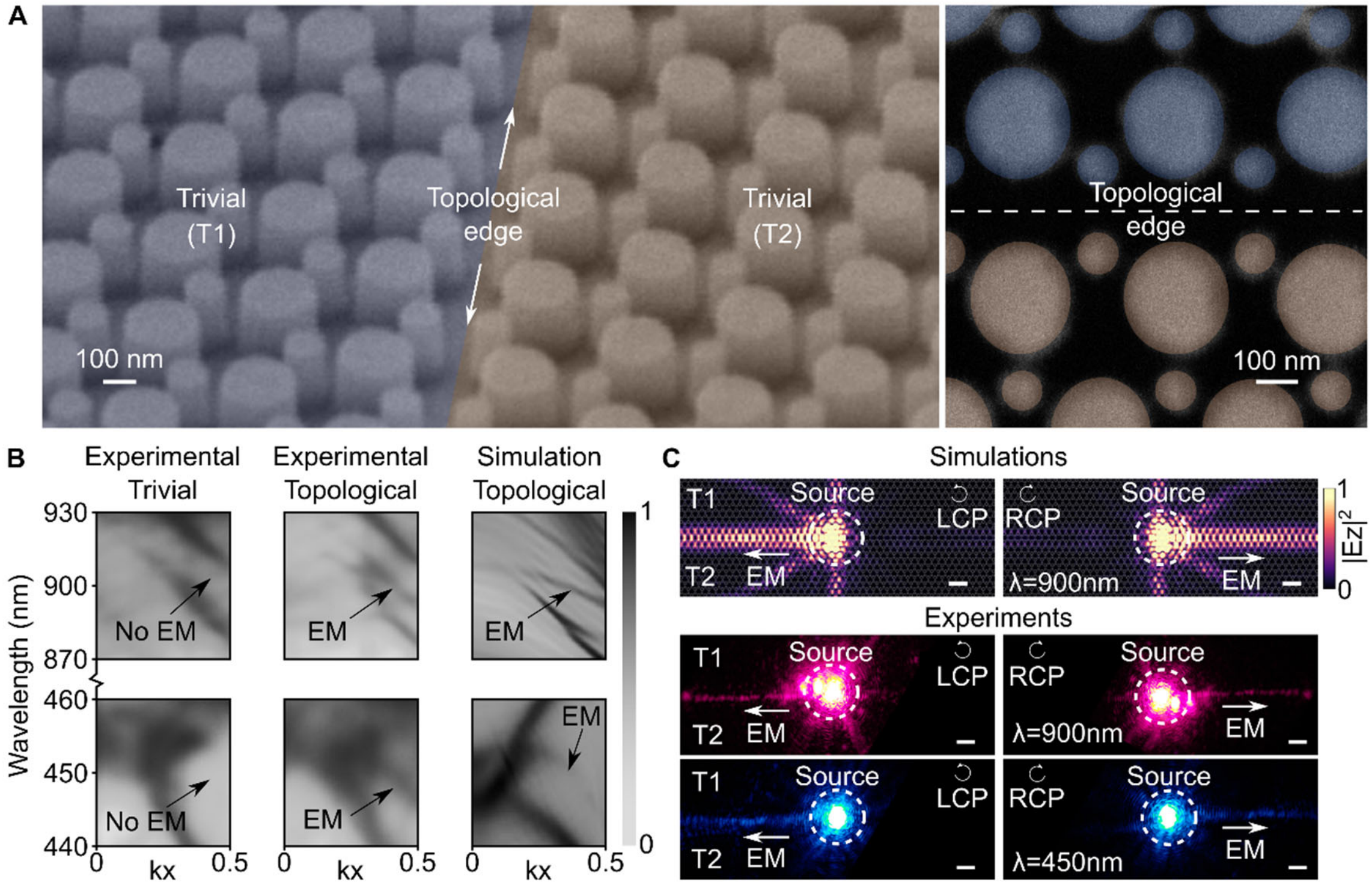

**Fig. 2. Linear topological modes observed at two frequencies.**
**(A)** Scanning electron microscope (SEM) image of the fabricated HPPTI showing a topological edge separating two domains (false-colored) with different topologies. Note that within a domain or without an edge is not topological or so-called trivial. The right panel shows a magnified top view of the topological edge, and the scale bars represent 100 nm. **(B)** Wavevector-resolved reflection spectrum under a broadband illumination distinguishes trivial bulk and topological edge responses. Experimental measurements from a trivial region (left) show no edge modes (EMs), while spectra acquired at the topological edge (middle) reveal two EMs near wavelengths of 900 nm and 450 nm. Full-wave simulations (right) at the topological edge validate the presence of EMs. **(C)** Chiral edge-mode propagation under circularly polarized excitation. Simulations (top) and experiments (bottom) show that left- and right-circularly polarized (LCP and RCP) excitation launches counter-propagating EMs. Reversing helicity switches the propagation direction, confirming spin-orbit locking in the topological edge states. The simulations at the second harmonic wavelength, 450 nm, are discussed in supplementary note 5, since they exhibit similar spin-momentum locking behavior. Scale bars represent 5 µm.

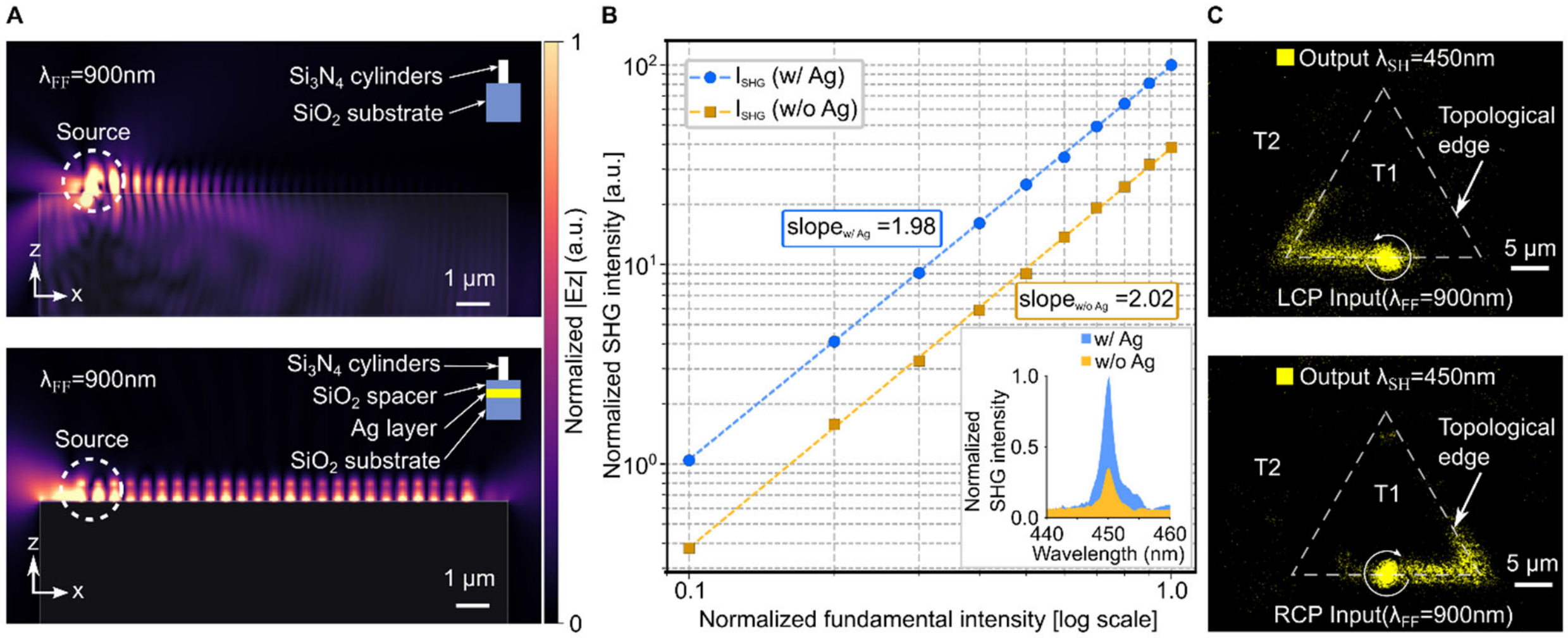

**Fig. 3. Nonlinear topological states empowered by surface plasmonics.**
**(A)** Simulated out-of-plane electric field magnitude |Ez| at $\lambda_{FF}$ = 900 nm for structures without (top) and with (bottom) a silver back reflector. Without silver, the optical field leaks substantially into the substrate. With the silver back reflector separated from the active $Si_3N_4$ layer by a thin $SiO_2$ spacer, the field is strongly within the high-index (~2) silicon nitride region, suppressing substrate leakage and increasing the local field amplitude at the nonlinear interface. **(B)** Log-log plot of normalized SHG intensity at $\lambda_{SH}$ = 450 nm versus fundamental input power at $\lambda_{FF}$ = 900 nm for devices with and without the silver layer. Both exhibit near-quadratic scaling with slopes of 1.98 and 2.02, respectively, confirming second-order nonlinear behavior, while the metal-backed structure delivers a consistent threefold enhancement in SHG intensity across the full power range. Inset: SHG spectra showing a pronounced peak at $\lambda_{SH}$ = 450 nm, with the silver-backed device producing a significantly stronger and spectrally clean second harmonic signal. **(C)** Chiral steering of topologically protected SHG edge modes along a sharply bent topological edge. Under LCP and RCP excitation, the generated SHG remains confined to the triangular edge and propagates in opposite directions, demonstrating polarization-controlled chiral transport with spin-momentum locking.

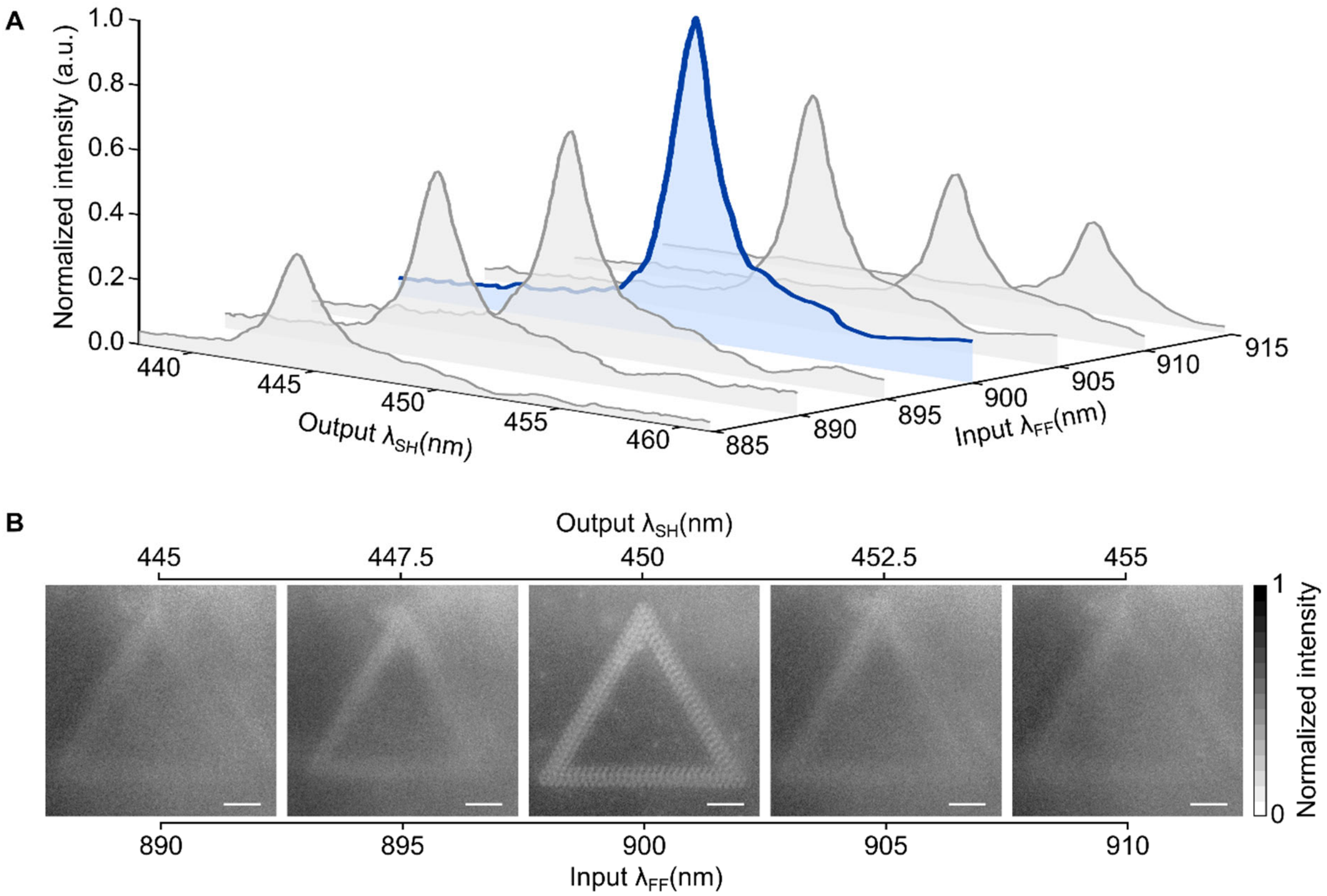

**Fig. 4. Enhanced frequency conversion efficiency through phase matching.**

**(A)** Experimentally acquired nonlinear intensity as a function of wavelength at fundamental frequency ($\lambda_{FF}$) and second harmonic ($\lambda_{SH}$) for a triangular topological edge. As the $\lambda_{FF}$ is tuned (885–915 nm), an intensity peak is observed precisely at the doubled frequency or the half wavelength, verifying the existence of $\chi^{(2)}$ in the PECVD-deposited $Si_3N_4$ material. The intensity peak reaches the maximum at $\lambda_{FF}$ = 900 nm (blue shaded), indicating that the phase-matching condition is fulfilled and the two topological modes are interconnected. Such wavelength-dependent SHG is widely used for classical nonlinear phase matching. The SHG spectra is obtained by integrating the emitted intensity over all pixels within the region of interest, while maintaining a constant pump power across all excitation wavelengths. **(B)** Real-space images of SHG at selected $\lambda_{FF}$ (890-910 nm). Besides maximum SHG intensity, a strong confinement is observed at $\lambda_{FF}$ = 900 nm, where phase matching is satisfied, while other wavelengths exhibit reduced and delocalized SHG. These results with delocalization and spatial confinement of SHG while keeping the beam size of the pump intact demonstrate the bulk nature of $\chi^{(2)}$ rather than originating from surfaces. More importantly, they also show that phase matching selectively enhances and localizes the SHG at the topological edge following the triangular geometry.

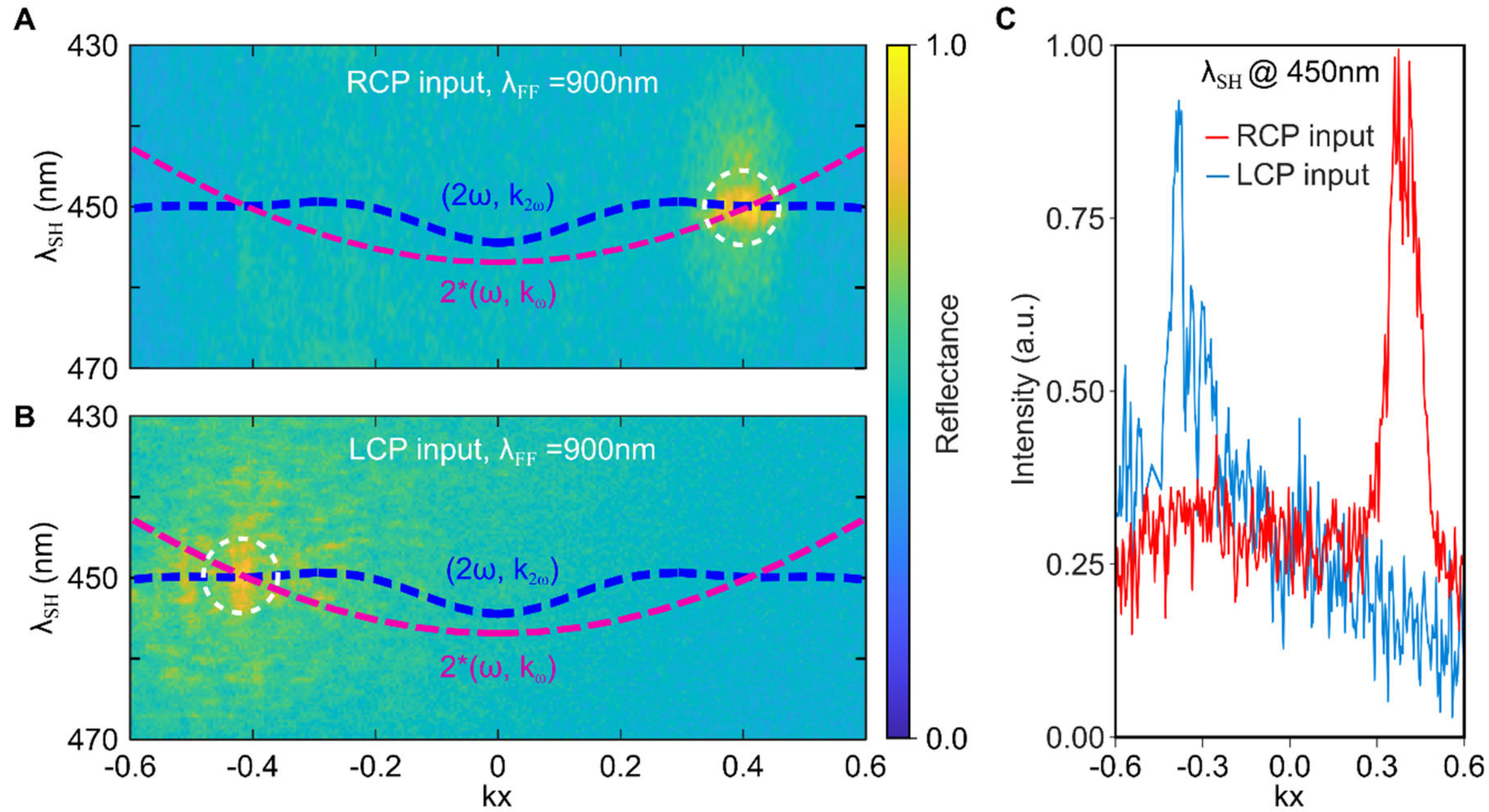

**Fig. 5. Phase matching of double topological states with spin-momentum locking.**
**(A), (B)** Wavevector-resolved reflection spectra measured under RCP and LCP fundamental excitations at ω ($\lambda_{FF}$ = 900 nm), with the detected signal corresponding to second harmonic emissions at 2ω ($\lambda_{SH}$ = 450 nm). Overlaid red and blue dashed curves indicate the simulated edge modes at 2*(ω, $k_\omega$) and (2ω, $k_{2\omega}$), respectively, highlighting nonlinear phase matching at the crosspoints with $k_{2\omega} = 2k_\omega$ (circles). In stark contrast to classical nonlinear phase matching, the intensity spots near $k_x = 0.4$ and $k_x = -0.4$ for RCP and LCP fundamental pump demonstrate a unique spin-momentum locking between the phase-matched topological states at ω and 2ω, where the helicity of the pump uniquely determines the propagation direction of the nonlinear output. **(C)** Line cut along $\lambda_{SH}$ = 450 nm extracted from the reflection spectra, showing the normalized SHG intensity as a function of in-plane wavevector for RCP (red) and LCP (blue) excitation. The asymmetric peaks confirm the spin-momentum locking with a helical nature of the interaction between the two topological states.